\documentclass[aps,pra,10pt,twocolumn,amsmath,amssymb,
    longbibliography,superscriptaddress,nofootinbib]{revtex4-1}

\usepackage[colorlinks=true,citecolor=blue,linkcolor=blue]{hyperref}
\usepackage{graphicx}
\usepackage[dvipsnames]{xcolor}
\usepackage{amsmath,amsthm,amssymb}
\usepackage{bm,bbm,mathbbol}
\usepackage{oubraces}

\newcommand{\eq}[1]{Eq.~(\ref{eq:#1})}
\newcommand{\eqs}[2]{Eqs.~(\ref{eq:#1}) and~(\ref{eq:#2})}
\newcommand{\equ}[1]{Equation~(\ref{eq:#1})}
\newcommand{\equs}[2]{Equations~(\ref{eq:#1}) and (\ref{eq:#2})}
\newcommand{\fig}[1]{Fig.~\ref{fig:#1}}
\newcommand{\sect}[1]{Sec.~\ref{sec:#1}}

\def\beq{\begin{equation}}
\def\eeq{\end{equation}}
\def\bea{\begin{eqnarray}}
\def\eea{\end{eqnarray}}
\def\nn{\nonumber\\}

\def\ket#1{\vert#1\rangle}
\def\bra#1{\langle#1\vert}
\def\ip#1#2{\langle#1\vert#2\rangle}
\def\me#1#2#3{\langle#1\vert#2\vert#3\rangle}
\def\wt#1{\widetilde{#1}}

\def\Re{{\rm Re\,}}
\def\Im{{\rm Im\,}}
\def\Tr{{\rm Tr\,}}

\def\kk{{\bm k}}
\def\rr{{\bm r}}
\def\vv{{\bm v}}
\def\Aa{{\bm A}}
\def\RR{{\bm R}}

\def\KK{{\bm K}}
\def\MM{{\bm M}}
\def\mm{{\bm m}}
\def\PP{{\bm P}}
\def\E{{\mathcal E}}
\def\EE{{\bm{\mathcal E}}}
\def\jj{{\bm j}}

\def\nhat{\hat{\bm n}}
\def\one{\hat{\mathbbm{1}}}
\def\aiso{\alpha_{\rm iso}}
\def\eps{\epsilon}
\def\alfa{{\mathfrak a}}

\begin{document}

\title{Geometric and nongeometric contributions to the surface
  anomalous Hall conductivity }

\author{Tom\'{a}\v{s} Rauch} \affiliation{Centro de F{\'i}sica de
  Materiales, Universidad del Pa{\'i}s Vasco (UPV/EHU), 20018 San
  Sebasti{\'a}n, Spain}

\author{Thomas Olsen} \affiliation{CAMD, Department of Physics,
  Technical University of Denmark, 2820 Kgs. Lyngby Denmark}

\author{David Vanderbilt} \affiliation{Department of Physics and
  Astronomy, Rutgers University, Piscataway, New Jersey 08854-8019,
  USA}

\author{Ivo Souza} \affiliation{Centro de F{\'i}sica de Materiales,
  Universidad del Pa{\'i}s Vasco (UPV/EHU), 20018 San Sebasti{\'a}n,
  Spain} \affiliation{Ikerbasque Foundation, 48013 Bilbao, Spain}

\date{\today}

\begin{abstract}
  A static electric field generates circulating currents at the
  surfaces of a magnetoelectric insulator. The anomalous Hall part of
  the surface conductivity tensor describing such bound currents can
  change by multiples of $e^2/h$ depending on the insulating surface
  preparation, and a bulk calculation does not fix its quantized
  part. To resolve this ambiguity, we develop a formalism for
  calculating the full surface anomalous Hall conductivity in a slab
  geometry. We identify a Berry-curvature term, closely related to the
  expression for the bulk anomalous Hall conductivity, whose value can
  change by quantized amounts by adjusting the surface Hamiltonian.
  In addition, the surface anomalous Hall conductivity contains a
  nongeometric part that does not depend on the surface preparation.
\end{abstract}
\maketitle

\section{Introduction}
\label{sec:intro}

Certain surface properties of crystals are strongly constrained by the
bulk, and as a result they are very robust with respect to local
perturbations. An example is the areal charge density
$\sigma_{\rm surf}$ bound to an insulating surface of a polar
insulator.  For an unreconstructed defect-free surface with outward
normal $\nhat$ it is given by~\cite{vanderbilt-prb93}
\beq
\label{eq:sigma-P}
\sigma_{\rm surf}=
\left(\PP+\frac{e\RR}{V_{\rm c}}\right)\cdot\nhat,
\eeq
where ${\bm P}$ is the bulk electric polarization, $\RR$ is a lattice
vector, and $V_c$ is the volume of a unit cell. According to the
Berry-phase theory~\cite{king-smith-prb93}, $\PP$ is only defined
modulo $e\RR/V_{\rm c}$, since it is possible to change its value by
that amount by adjusting the phases of the Bloch wave functions.
\equ{sigma-P} assumes that a definite choice of gauge has been made so
that a unique value of $\PP$ has been established. (Here, the word
``gauge'' refers to the freedom to adjust the phases of the Bloch
eigenstates or, more generally, to perform a unitary transformation at
each $\kk$ among the occupied Bloch states~\cite{marzari-prb97}.) The
second term in \eq{sigma-P} amounts to an integer number of electrons
per surface unit cell. Its presence is required because it is in
principle possible to prepare the insulating surface in different ways
such that the macroscopic charge per surface cell changes by a
multiple of the elementary charge~$e$.  Thus, the quantized part of
$\sigma_{\rm surf}$ depends on the details at the surface but the
nonquantized part does not.

In this work, we consider a similar situation that arises in
insulating crystals that display the linear magnetoelectric (ME)
effect, whereby an applied magnetic field~$\bm{B}$ induces an electric
polarization $\PP$, and conversely an applied electric field $\EE$
induces a magnetization $\MM$~\cite{landau-EM,fiebig-jpd05}. The
linear ME tensor is defined as
\beq
\label{eq:alpha}
\alpha_{ab}=\left.\frac{\partial P_a}{\partial B_b}\right|_{\EE=0}
=\left.\frac{\partial M_b}{\partial \E_a}\right|_{{\bm B}=0}.
\eeq
The full ME response contains both frozen-ion and lattice-mediated
contributions, and each can be further decomposed into spin and
orbital parts. In the following, we focus exclusively on the
frozen-ion orbital response.

The bulk magnetization generated by a static electric field gives rise
to surface currents
\beq
\label{eq:K}
 \KK=\MM\times\nhat.
\eeq
In the case of an insulating surface, this is the full current
response.  It is described at linear order by a $2\times 3$ surface
conductivity tensor
$\sigma^{\rm surf}_{ab}=\partial K_a/\partial \E_b$, and the surface
anomalous Hall conductivity (AHC) is defined as the antisymmetric part
of the $2\times 2$ block that describes the surface current generated
by an in-plane electric field.  Writing the surface AHC in vector form
as $\sigma^{\rm AH}_{\rm surf}\,\nhat$ where
\beq
\label{eq:sigma-surf}
\sigma^{\rm AH}_{\rm surf}
=-\frac{1}{2}\eps_{cdb}\sigma^{\rm surf}_{cd}\hat n_b,
\eeq
the surface anomalous Hall current density becomes
\beq
\label{eq:K-AH}
\KK^{\rm AH}=\sigma^{\rm AH}_{\rm surf}\,\nhat\times\EE.
\eeq
From \eqs{alpha}{K} we find
\beq
\sigma^{\rm surf}_{cd} = \frac{\partial K_c}{\partial \E_d}  
=  \frac{\partial}{\partial \E_d}\eps_{cea}M_e \hat n_a 
= \eps_{cea}\alpha_{de} \hat n_a,
\eeq
and plugging this expression into \eq{sigma-surf} leads to
\beq
\label{eq:sigma-AH-surf}
\sigma^{\rm AH}_{\rm surf}
 = -\frac{1}{2} {\rm Tr}(\alpha)+\frac{1}{2} \alpha_{ab} \hat n_a \hat n_b. 
\eeq
Separating the ME tensor on the right-hand side into an isotropic
trace piece and a traceless part,
\beq
\label{eq:alpha-trace-traceless}
\alpha_{ab}=\aiso\delta_{ab}+\wt\alpha_{ab},
\eeq
we arrive at the relation
\beq
\label{eq:sigma-alpha-a}
\sigma^{\rm AH}_{\rm surf}:=-\aiso+
\frac{1}{2}\wt\alpha_{ab}\hat n_a\hat n_b.
\eeq
We use the special symbol := to indicate that while the left-hand side
is uniquely defined for a given surface termination, the right-hand
side carries a quantum of indeterminacy, since the bulk quantity
$\aiso$ is gauge invariant only modulo
$e^2/h$~\cite{qi-prb08,essin-prl09}.

Once a definite value has been chosen for the multivalued quantity
$\aiso$, \eq{sigma-alpha-a} can be rewritten as
\beq
\label{eq:sigma-alpha}
\sigma^{\rm AH}_{\rm surf}=-\aiso+m\frac{e^2}{h}+
\frac{1}{2}\wt\alpha_{ab}\hat n_a\hat n_b.
\eeq
Different choices of the integer $m$ in the middle term correspond to
different surface preparations exhibiting values of the surface AHC
that differ by a multiple of the quantum of conductance.%
\footnote{There are two scenarios compatible with \eq{sigma-alpha}. If
  the integer $m$ is the same for all crystal facets, the entire
  surface is insulating and the term $me^2/h$ gives an isotropic
  contribution to the surface AHC~\cite{coh-prb11}.  If adjacent
  facets have different $m$ values, there are chiral conducting
  channels along the connecting hinges~\cite{sitte-prl12}.}  This
integer can be changed, in principle, by stitching a quantum anomalous
Hall layer to the surface~\cite{essin-prl09,coh-prb11} or otherwise
changing the surface Hamiltonian, or by means of an adiabatic pumping
cycle characterized by a nonzero second Chern
number~\cite{taherinejad-prl15,olsen-prb17b}. It follows that only the
nonquantized part of the surface AHC is a bulk property, in close
analogy with \eq{sigma-P} for the surface charge. These features are
described by the phenomenology of axion
electrodynamics~\cite{qi-prb08,wilczek-prl87}, and $\aiso$ is
sometimes referred to as the axion ME coupling.

We are now ready to formulate the main question behind this work.
Suppose we have a ME insulator (it should break both inversion and
time reversal symmetry), and we consider a specific insulating
surface.  How can we calculate the surface AHC, not just up to a
quantum but exactly?  Since we are given a definite surface
Hamiltonian, there should be a definite answer without any quantum of
ambiguity. We shall answer this question by developing a formalism
that allows one to calculate the surface AHC unambiguoulsy using a
slab geometry.

The paper is organized as follows.  We begin in \sect{ahc-local} by
calculating, at linear order, the local current response of an
insulator to a static homogeneous electric field.  The local AHC,
defined as the antisymmetric part of this local conductivity tensor,
is then separated into geometric and nongeometric parts.  Starting
from the expression for the local AHC, we obtain in \sect{ahc-surf} an
expression for the surface AHC of a slab, which we again separate into
geometric and nongeometric parts.  In \sect{numerical} we calculate
numerically the surface AHC for slabs of tight-binding (TB) models and
compare the results, via \eq{sigma-alpha}, with independent
calculations of the bulk ME tensor. We conclude in \sect{summary} with
a summary, and leave a lengthier derivation to an appendix.

\section{Local anomalous Hall conductivity}
\label{sec:ahc-local}

\subsection{Linear-response calculation}
\label{sec:linres}

The local conductivity and local AHC are defined as
\beq
\label{eq:sigma-loc}
\sigma_{ab}(\rr)=
\left.\frac{\partial j_a(\rr)}{\partial \E_b}\right|_{\EE=0}
\eeq
and
\beq 
\label{eq:ahc-vec}
\sigma^{\rm AH}_c(\rr)=-\frac{1}{2}\eps_{abc}\sigma_{ab}(\rr),
\eeq
respectively.  $\EE$ denotes a static homogeneous electric field and
$\jj(\rr)$ is the microscopic induced current density, whose anomalous
Hall part is given by
$\jj^{\rm AH}(\rr)={\bm \sigma}^{\rm AH}(\rr)\times\EE$.

We wish to calculate the local AHC for an insulating medium at zero
temperature described by a single-particle Hamiltonian $\hat H$. The
current operator is
\beq
\label{eq:j-op}
\hat\jj(\rr)=-\frac{e}{2}
\left(\ket{\rr}\bra{\rr}\hat\vv+\hat\vv\ket{\rr}\bra{\rr}\right),
\eeq
where $\hat\vv=(1/i\hbar)[\hat\rr,\hat H]$ is the velocity operator
and $e>0$ is the elementary charge. The current density is given by
\beq
\label{eq:j}
\jj(\rr)=\Tr\left[\hat P\hat\jj(\rr)\right]
=-e\,\Re\me{\rr}{\hat \vv\hat P}{\rr},
\eeq
where $\hat P$ denotes the projection operator onto the occupied
states. An expression for the local conductivity~(\ref{eq:sigma-loc})
can now be obtained by differentiating \eq{j} with respect to $\EE$.
For that purpose we write $\hat H=\hat H_0+e\EE\cdot\hat\rr$ where
$\hat H_0$ is the unperturbed Hamiltonian, and note that since
$[\hat r_a,\hat r_b]=0$ the operator $ \hat\vv$ reduces to
$(1/i\hbar)[\rr,\hat H_0]$.  Hence the electric field enters \eq{j}
via $\hat P$ only,\footnote{We are ignoring local-field corrections,
  which introduce a dependence of $\hat H_0$ on $\EE$ through the
  self-consistent charge density. Such terms are not difficult to
  derive, but they are absent from our non-self-consistent TB
  calculations.}  leading to
$\sigma_{ab}(\rr) = (-e)\Re\me{\rr}{\hat v_a\partial_{\E_b}\hat
  P}{\rr}$, and inserting this expression in \eq{ahc-vec} we arrive at
\beq
\label{eq:ahc-r} 
{\bm\sigma}^{\rm AH}(\rr) =\frac{e}{2}\,
\Re\me{\rr}{\hat\vv\times\partial_\EE\hat P}{\rr}.
\eeq
Finally, from first-order perturbation theory we get
\beq
\label{eq:dPdE}
\partial_\EE\hat P=-e\sum_{v,c}\,
\left(
  \ket{c}\frac{\me{c}{\hat\rr}{v}}{E_{cv}}\bra{v}+
  \ket{v}\frac{\me{v}{\hat\rr}{c}}{E_{cv}}\bra{c}
\right),
\eeq
where $\ket{v}$ and $\ket{c}$ denote occupied and empty energy
eigenstates, respectively, and $E_{cv}=E_c-E_v$. \equs{ahc-r}{dPdE}
give the full local AHC; below, we separate it into geometric and
nongeometric parts.

\subsection{Separation of the local AHC into geometric and
  nongeometric parts}
\label{sec:separation-local}

Consider the isotropic ME response of a bounded sample, defined as
\beq
\label{eq:aiso-tilde}
\alfa_{\rm iso}
=\frac{1}{3}\sum_{a=1}^3\, \frac{\partial m_a}{\partial\E_a}
\eeq
in terms of the orbital moment
\beq
\label{eq:m}
{\bm m}=\frac{1}{2}\int \rr\times{\bm j}(\rr)\,d^3r.
\eeq
For a globally insulating crystallite of volume $V$,
$\alfa_{\rm iso}/V$ converges in the $V\rightarrow\infty$ limit to one
of the multiple values of $\aiso$, with the specific value depending
on the surface preparation~\cite{coh-prb11}. Plugging \eq{m} into
\eq{aiso-tilde} and comparing with the definition of the local AHC in
\eq{ahc-vec}, we find
\beq
\label{eq:alfa-sigma}
\alfa_{\rm iso} = -\,\frac{1}{3}\,
  \int \rr\cdot\bm{\sigma}^{\rm AH}(\rr)\, d^3r.
\eeq
This relation will be used below to isolate the geometric and
nongeometric contributions to the local AHC, but first we need some
results from the microscopic theory of the orbital ME response in
insulators~\cite{malashevich-njp10,essin-prb10}.

The quantum-mechanical expression for the bulk ME tensor $\alpha_{ab}$
comprises an isotropic geometric term known as the Chern-Simons (CS)
term, and a nongeometric term known as the Kubo or cross-gap (cg) term
that has both isotropic and anisotropic
parts~\cite{malashevich-njp10,essin-prb10}.  The relation between
those two terms and the decomposition in \eq{alpha-trace-traceless}
can be summarized as follows:
\beq
\label{eq:alpha-decomp}
\alpha_{ab}=
\overunderbraces{&&\br{2}{\alpha^{\text{cg}}_{ab}}}
{&\big(\alpha_{\rm CS}+&\aiso^{\text{cg}}\big)
&\delta_{ab}+\wt\alpha_{ab}}{&\br{2}{\aiso}}.
\eeq
The expressions for $\alpha_{\rm CS}$ and $\alpha^{\text{cg}}_{ab}$
take the form of integrals over the Brillouin zone (BZ), and can be
found in Refs.~\onlinecite{malashevich-njp10}
and~\onlinecite{essin-prb10}. In the case of $\alpha_{\rm CS}$, the
integrand only contains the unperturbed cell-periodic Bloch functions
and their first $\kk$ derivatives; it is a gauge-dependent quantity,
but after integration over the entire BZ it becomes gauge invariant
modulo $e^2/h$. The expression for $\alpha^{\text{cg}}_{ab}$ contains
in addition the perturbed wave functions and the velocity operator,
and is fully gauge invariant at each point in the BZ.

The ME tensor $\alfa_{ab}=(\partial m_b/\partial\E_a)_{{\bm B}=0}$ of
a finite crystallite can be similarly decomposed into geometric and
nongeometric terms~\cite{malashevich-njp10}.  Because surface
contributions are now included, the geometric part $\alfa_{\rm CS}$ of
the isotropic piece $\alfa_{\rm iso}$ is unique for a given surface
preparation. It is given by~\cite{malashevich-njp10}
\bea
\label{eq:alpha-iso-cs}
\alfa_{\rm CS}&=&-\frac{2\pi e^2}{3h}\eps_{abc}
\Im\Tr\left[\hat P_0\hat r_a\hat P_0\hat r_b\hat P_0\hat r_c\right]\nn
&=&\frac{2\pi e^2}{3h}\eps_{abc}
\int r_c\,\Im\me{\rr}{\hat P_0\hat r_a\hat Q_0\hat r_b\hat P_0}{\rr}\,
d^3r,
\eea
where $\hat P_0=\sum_v\ket{v}\bra{v}$ and $\hat Q_0=\one-\hat P_0$ are
the ground-state projector and its complement, respectively. Comparing
with \eq{alfa-sigma}, we are led to identify a geometric (CS)
contribution to the local AHC given by
\begin{subequations}
\label{eq:ahc-CS-r}
\begin{align}
\label{eq:ahc-CS-r-a}
{\bm\sigma}^{\rm AH}_{\rm CS}(\rr)&=\frac{e^2}{h}{\bm C}(\rr),\\
\label{eq:C-r}
{\bm C}(\rr)&=-2\pi\Im\me{\rr}
{\hat P_0\hat\rr\hat Q_0\times\hat Q_0\hat\rr\hat P_0}{\rr}.
\end{align}
\end{subequations}

One can obtain the nongeometric (cross-gap) part of the local AHC in a
similar way, starting from the nongeometric part of the orbital ME
coupling of a crystallite.  This is done in the Appendix and as
expected the result is that the cross-gap local AHC is equal to the
difference between the full local AHC~(\ref{eq:ahc-r}) and the CS term
above,
\beq
\label{eq:ahc-cg-r}
{\bm\sigma}^{\rm AH}_{\rm cg}(\rr)=
{\bm\sigma}^{\rm AH}(\rr)-{\bm\sigma}^{\rm AH}_{\rm CS}(\rr).
\eeq

Equations~(\ref{eq:ahc-r}), (\ref{eq:ahc-CS-r}),
and~(\ref{eq:ahc-cg-r}) are the main results of this section.

\subsection{Discussion}
\label{sec:discussion-local}

The appearance of a nongeometric term in the local AHC may seem
surprising at first, since the intrinsic macroscopic AHC of a bulk
crystal or slab is known to be purely geometric: it is given by the BZ
integral of the Berry curvature of the occupied Bloch
states~\cite{xiao-rmp10}. The explanation, as we will see
\sect{discussion-surface}, is that the nongeometric term always
integrates to zero across the full width of a slab, dropping out from
the net AHC of the slab.  As will become clear in the following, that
term does contribute to the AHC of a single surface.

The nongeometric part of the local AHC was overlooked in some previous
studies~\cite{essin-prl09,marrazzo-prb17}, where the local AHC was
formulated as a spatially-resolved Berry curvature. As for the
geometric part, the expression in \eq{ahc-CS-r} is consistent with the
previous literature.  Consider a flat crystallite lying on the $(x,y)$
plane, and integrate the quantity $C_z(\rr)$ given by \eq{C-r} over
all $z$ to obtain a dimensionless quantity $C_z(x,y)$. This is
precisely the ``local Chern number'' introduced in
Ref.~\cite{bianco-prb11}.  For a slab, the average of \eq{ahc-CS-r-a}
over a two-dimensional (2D) cell at fixed $z$ is essentially identical
to the ``layer-resolved AHC'' of Ref.~\cite{essin-prl09}.

In the next section we calculate the layer-resolved AHC including both
geometric and nongeometric contributions, and use it to evaluate the
surface AHC.

\section{Surface anomalous Hall conductivity}
\label{sec:ahc-surf}

\subsection{Evaluation in a slab geometry}
\label{sec:ahc-surf-slab}

Consider an insulating slab with the outward normal
$\nhat=\hat{\bm z}$ of the top surface pointing along a
reciprocal-lattice vector ${\bm b}_3$. We assume that the slab
thickness $L$ is much larger than the lattice
constant~$c=2\pi/|{\bm b}_3|$ in the surface-normal direction. We also
assume a defect-free surface, and introduce a ``layer-resolved'' AHC
for the slab by averaging the $z$ component of the local
AHC~(\ref{eq:ahc-vec}) over a surface unit cell at fixed~$z$:
\beq
\label{eq:ahc-z-def}
\sigma^{\rm AH}_{\rm slab}(z)
=\frac{1}{A_c}\int_{A_c}\sigma^{\rm AH}_z(x,y,z)\,dx\,dy.
\eeq

The net AHC of a slab is given by
\beq
\label{eq:ahc-slab-def}\sigma^{\rm AH}_{\rm slab}
=\int\sigma^{\rm AH}_{\rm slab}(z)\,dz, 
\eeq
where the range of integration is chosen to span the full width of the
slab, including the exponential tails of the wave functions outside
the two surface regions.  For an insulating slab the result is
quantized in units of $e^2/h$~\cite{xiao-rmp10}:
\beq
\label{eq:ahc-slab}
\sigma^{\rm AH}_{\rm slab}=\frac{e^2}{h}C_{\rm slab},
\eeq
where the integer $C_{\rm slab}$ is the Chern number of the slab (see
\sect{discussion-surface}). In order for this equation to be
meaningful, we are assuming that the slab is cut from a bulk insulator
for which all three of the bulk Chern indices
$C_j$~\cite{kohmoto-pb93} are zero.

As a first step towards calculating the surface AHC, we filter out the
atomic-scale oscillations in the layer-resolved AHC by performing a
``sliding-window average'' over one vertical lattice constant:
\beq
\label{eq:ahc-z-avg}
\overline\sigma^{\rm AH}_{\rm slab}(z)=\frac{1}{c}
\int_{z-c/2}^{z+c/2}\sigma^{\rm AH}_{\rm slab}(z')\,dz'.
\eeq
Because we assumed $C_3=0$, this coarse-grained AHC must vanish
exponentially in the bulklike interior region of the slab, and it can
only become nonzero near the two surfaces.  The macroscopic AHC of the
top surface can now be expressed as
\beq
\label{eq:ahc-surf-def}
\sigma^{\rm AH}_{\rm surf}=\int_{z_0}
\overline\sigma^{\rm AH}_{\rm slab}(z)\,dz,
\eeq
with $z_0$ chosen in the bulklike region of the slab, and the upper
limit of integration placed at an arbitrary point in the vacuum region
above the top surface.  The AHC of the bottom surface is
$(e^2/h)C_{\rm slab}-\sigma^{\rm AH}_{\rm surf}$.

For numerical work, it is more convenient to recast \eq{ahc-surf-def}
as
\beq
\label{eq:ahc-surf-practical}
\sigma^{\rm AH}_{\rm surf}=
\int\sigma^{\rm AH}_{\rm slab}(z)f_{\rm ramp}(z-z_0)\,dz,
\eeq
where the range of integration spans the full width of the slab, and
$f_{\rm ramp}$ is a ramp-up function defined as
\beq
\label{eq:f}
f_{\rm ramp}(z)=
\begin{cases}
0,\text{ for }z<-c/2\\
z/c+1/2,
\text{ for }-c/2<z<c/2\\
1,\text{ for }z>c/2.
\end{cases}
\eeq

To summarize, \eq{ahc-surf-practical} gives the surface AHC in terms
of the layer-resolved AHC of \eq{ahc-z-def}, for which we provide an
explicit formula below.
\\

\subsection{Layer-resolved anomalous Hall conductivity}
\label{sec:ahc-surf-tot}

We evaluate the layer-resolved AHC by inserting \eq{ahc-r} for the
local AHC into \eq{ahc-z-def}. The ground-state projector expressed in
terms of the valence eigenstates of the slab reads as
\beq
\label{eq:P-slab}
\hat P_0
=\frac{1}{N}\sum_{\kk v}\,\ket{\psi_{\kk v}}\bra{\psi_{\kk v}}
\eeq
(the summation in $\kk=(k_x,k_y)$ is over a uniform mesh of $N$ points
covering the surface BZ), and we need its linear change under an
in-plane electric field,
\beq
\label{eq:dPdE-slab}
\partial_\EE\hat P=\frac{1}{N}\sum_{\kk v}\,
e^{i\kk\cdot\hat\rr}
\left(
  \ket{\wt\partial_\EE u_{\kk v}}\bra{u_{\kk v}}+
  \ket{u_{\kk v}}\bra{\wt\partial u_{\kk v}}
\right)
e^{-i\kk\cdot\hat\rr}.
\eeq
Here, $\ket{u_{\kk v}}$ is the cell-periodic part of
$\ket{\psi_{\kk v}}$, and
\beq
\label{eq:dudE}
\ket{\wt{\bm\partial}_\EE u_{\kk v}}=ie\sum_c\,\ket{u_{\kk c}}\,
\frac{\hbar \vv_{\kk cv}}{E_{\kk cv}^2}
\eeq
is the projection of $\ket{{\bm\partial_\EE u_{\kk v}}}$ onto the
conduction bands, with
$\vv_{\kk cv}=\me{u_{\kk c}}{\hat \vv_\kk}{u_{\kk v}}$ and
$\hat\vv_\kk =e^{-i\kk\cdot\hat\rr}\hat\vv e^{i\kk\cdot\hat\rr}$.
\equ{dPdE-slab} is essentially the same as \eq{dPdE}, but adapted to a
slab geometry.  Putting everything together and letting
$N\rightarrow\infty$, we arrive at
\begin{widetext}
\beq
\label{eq:ahc-z}
\sigma^{\rm AH}_{\rm slab}(z)
=\frac{e}{4\pi h}\int d^2k
\int_{A_c}dx\,dy\,\sum_v\,\Re
\left[
  \bra{\rr} 
  \hbar \hat\vv_\kk\times
  \left(
    \ket{\wt{\bm\partial}_\EE u_{\kk v}}\ip{u_{\kk v}}{\rr}+
    \ket{u_{\kk v}}\ip{\wt{\bm\partial}_\EE u_{\kk v}}{\rr}
  \right)
\right]_z,
\eeq
\end{widetext}
where the integral in $\kk$ is over the surface BZ.

\subsection{Separation of the layer-resolved AHC into geometric and
  nongeometric parts}
\label{sec:ahc-surf-cs}

To find the geometric part of the layer-resolved AHC, we repeat the
above steps, simply replacing \eq{ahc-r} for the full local AHC with
\eq{ahc-CS-r} for the geometric part.  Inserting \eq{P-slab} for
$\hat P_0$ in \eq{C-r} gives
\beq
C_z(\rr)=-\frac{4\pi}{N^2}\Im\sum_{\kk\kk'vv'}\,
\psi^*_{\kk v}(\rr)\psi_{\kk' v'}(\rr)
\me{\psi_{\kk v'}}{\hat x\hat Q_0\hat y}{\psi_{\kk v}}.
\eeq
Writing $\hat Q_0$ as
$(1/N)\sum_{\kk c}\,\ket{\psi_{\kk c}}\bra{\psi_{\kk c}}$ and using
the identity
$\me{\psi_{\kk v}}{\hat r_j}{\psi_{\kk' c}}= iN\ip{u_{\kk
    v}}{\partial_{k_j} u_{\kk c}}\delta_{\kk\kk'}$
valid for off-diagonal position matrix elements~\cite{blount-ssp62},
we obtain
\beq
\label{eq:C-r-slab}
C_z(\rr)
=-\frac{4\pi}{N}\Im\sum_\kk\,u^*_{\kk v}(\rr)u_{\kk v'}(\rr)
F_{\kk v'v}^{xy},
\eeq
where 
\beq
\label{eq:metric-curvature}
F_{\kk v'v}^{xy}
=\sum_c\,\ip{\partial_{k_x}u_{\kk v'}}{u_{\kk c}}
\ip{u_{\kk c}}{\partial_{k_y}u_{\kk v}}
=\left({\cal F}_{\kk vv'}^{yx}\right)^*
\eeq
is the metric-curvature tensor~\cite{marzari-prb97}.  Inserting
\eq{C-r-slab} in \eq{ahc-CS-r-a} for
${\bm\sigma}^{\rm AH}_{\rm CS}(\rr)$ and plugging the result into
\eq{ahc-z-def} for the layer-resolved AHC yields, for
$N\rightarrow\infty$,
\beq
\label{eq:ahc-z-cs-a}
\sigma^{\rm AH}_{\rm slab,CS}(z)=-\frac{e^2}{\pi h}\int 
\Im\Tr\left[{\cal O}_\kk(z)F^{xy}_\kk\right]\,d^2k.
\eeq
The trace is over the valence bands, and
\beq
\label{eq:O-z}
{\cal O}_{\kk vv'}(z)
=\int_{A_c}u^*_{\kk v}(x,y,z)u_{\kk v'}(x,y,z)\,dx\,dy
\eeq
is a layer-resolved overlap matrix.

\equ{ahc-z-cs-a} can be brought to a more transparent form by
decomposing the metric-curvature tensor into Hermitian and
anti-Hermitian parts in the band indices as
$F^{xy}_{\kk nm}=R^{xy}_{\kk nm}+(1/2i)\wt\Omega^{xy}_{\kk nm}$, where
\beq
 R^{xy}_{\kk nm}
=\frac{1}{2}F^{xy}_{\kk nm}+\frac{1}{2}\left(F^{xy}_{\kk mn}\right)^*
\eeq
is the quantum metric and
\beq
\label{eq:curv-cov-a}
\wt\Omega^{xy}_{\kk nm}=iF^{xy}_{\kk nm}+\left(iF^{xy}_{\kk mn}\right)^*
\eeq
is the gauge-covariant Berry curvature, related to the Berry
connection $A_{\kk nm}^a=i\ip{u_{\kk n}}{\partial_{k_a}u_{\kk m}}$ and
to the noncovariant Berry curvature
$\Omega^{xy}_{\kk nm}=\partial_{k_x}A^y_{\kk nm}
-\partial_{k_y}A^x_{\kk nm}$ by
\beq
\label{eq:curv-cov}
\wt\Omega^{xy}_{\kk nm}=\Omega^{xy}_{\kk nm}
-i\left[A^x_\kk,A^y_\kk\right]_{nm}.
\eeq
Since the matrices ${\cal O}_\kk(z)$, $R^{xy}_\kk$, and
$\wt\Omega^{xy}_\kk$ are all Hermitian and the trace of the product of
two Hermitian matrices is real, $R^{xy}_\kk$ drops out from
\eq{ahc-z-cs-a}, which reduces to
\beq
\label{eq:ahc-z-cs}
\sigma^{\rm AH}_{\rm slab,CS}(z)
=\frac{e^2}{2\pi h}
\int 
\Tr\left[{\cal O}_\kk(z)\wt\Omega^{xy}_\kk\right]\,d^2k.
\eeq
This expression for the CS layer-resolved AHC in terms of the
layer-resolved overlap matrix and the non-Abelian Berry curvature is
the central result of this section.

\equ{ahc-z-cs}, which essentially agrees with Eq.~(12) of
Ref.~\cite{essin-prl09},\footnote{\label{foot:PHC} To obtain the
  expression in Ref.~\cite{essin-prl09} for the CS layer-resolved AHC
  starting from the CS local AHC, one can repeat the derivation of
  \eq{ahc-z-cs} with a single modification: in \eq{C-r} for $C(\rr)$,
  exchange $\hat P_0$ and $\hat Q_0$ and remove the minus sign. It can
  be easily verified that this ``particle-hole transformation'' leaves
  $C(\rr)$ unchanged, as expected on physical grounds.} only accounts
for part of the layer-resolved AHC, whose full amount is given by
\eq{ahc-z}. According to \eq{ahc-cg-r} for the local AHC, the
remainder is the nongeometric (or cross-gap) part of the
layer-resolved AHC,
\beq
\label{eq:ahc-z-cross-gap}
\sigma^{\rm AH}_{\rm slab,cg}(z)= \sigma^{\rm AH}_{\rm
  slab}(z)-\sigma^{\rm AH}_{\rm slab,CS}(z).
\eeq

To review, the surface AHC is calculated from \eq{ahc-surf-practical},
where we insert either \eq{ahc-z} to obtain the grand total, or
\eq{ahc-z-cs} to find the geometric part.

\subsection{Discussion}
\label{sec:discussion-surface}

It was already mentioned in \sect{discussion-local} that although the
nongeometric part of the layer-resolved AHC can give a net
contribution to the surface AHC, its contribution to the AHC of the
entire slab vanishes.  To establish this result, let us show that the
geometric part of the slab AHC coincides with the total.

We begin with the total slab AHC. Inserting \eq{ahc-z} for the
layer-resolved AHC into \eq{ahc-slab-def} yields
\beq
\sigma^{\rm AH}_{\rm slab}
=-\frac{e^2}{2\pi h}\int d^2k\,\Im\sum_{vc}\,
\frac{\hbar^2 v^x_{\kk vc}v^y_{\kk cv}}
{E_{\kk cv}^2}-(x\leftrightarrow y).
\eeq
Using $\hbar\vv_{\kk vc}=-iE_{\kk cv}\Aa_{\kk vc}$ to remove the
energy denominator and noting that
$\Im\sum_{\kk vv'}A^x_{\kk vv'}A^y_{\kk v'v}=0$, the sum over
conduction bands~$c$ can be replaced by a sum over all bands $n$ (the
term $n=v$ vanishes). Comparing with the Berry curvature
$\Omega^{xy}_{\kk v}=-2\,\Im\sum_{n\not= v}\,A^x_{\kk vn}A^y_{\kk nv}$
we arrive at \eq{ahc-slab} for the total slab AHC, with the slab Chern
number given by~\cite{xiao-rmp10}
\beq
\label{eq:C-slab}
C_{\rm slab}=
\frac{1}{2\pi}\int d^2k\sum_v\,\Omega^{xy}_{\kk v}.
\eeq

To obtain the gometric part of the slab AHC, insert \eq{curv-cov} into
\eq{ahc-z-cs} and plug the result into \eq{ahc-slab-def}. Using
$\int{\cal O}_{\kk vv'}(z)\,dz =\delta_{vv'}$ we get
$\sigma^{\rm AH}_{\rm slab,CS}=C_{\rm slab}(e^2/h)$, which is the same
as the total slab AHC. Thus, \eq{ahc-z-cross-gap} must integrate to
zero across the entire slab.

The net amount of AHC contributed by the cross-gap
term~(\ref{eq:ahc-z-cross-gap}) across a single surface region is
equal to
\beq
\label{eq:ahc-surf-cg}
\sigma^{\rm AH}_{\rm surf,cg}=-\aiso^{\rm cg}
+\frac{1}{2}\wt\alpha_{ab}\hat n_a\hat n_b,
\eeq
and the CS term~(\ref{eq:ahc-z-cs}) contributes the additional amount
\beq
\label{eq:ahc-surf-cs}
\sigma^{\rm AH}_{\rm surf,CS}=-\alpha_{\rm CS}+m\frac{e^2}{h}.
\eeq
Together, they make up the full surface AHC of \eq{sigma-alpha}.

\section{Numerical results}
\label{sec:numerical}

In the following, we present the results of slab calculations of the
surface AHC for two different TB models, and compare the results with
bulk calculations of the orbital ME
tensor~\cite{malashevich-njp10}. First, let us briefly describe our TB
implementation of the surface AHC expressions.

\subsection{Tight-binding formulation of the surface AHC}
\label{sec:ahc-surf-TB}

In the TB context, the integration over~$z$ in \eq{ahc-surf-practical}
for the surface AHC gets replaced by a summation over a layer
index~$l$. The ramp-up function is evaluated at the discrete layer
coordinates $z(l)$, and the layer-resolved AHC becomes
$\sigma^{\rm AH}_{\rm slab}(l)$.

In \eq{ahc-z} for the layer-resolved AHC, $\ket{\rr}$ is replaced by
$\ket{i}$ representing a TB orbital $\phi_i(\rr)=\ip{\rr}{i}$, the
integration $\int_{A_c}dx\,dy$ is replaced by a summation
$\sum_{i\in l}$ over the orbitals within one surface unit cell in
layer $l$, and the matrix elements of the velocity operator are
evaluated by making the diagonal approximation
$\me{i}{\hat\rr}{j}={\bm\tau}_i\delta_{ij}$ for the position operator
in the TB basis~\cite{boykin-epj10}.

In \eq{ahc-z-cs} for the CS layer-resolved AHC the overlap matrix
becomes
\beq
 {\cal O}_{\kk vv'}(l)
=\sum_{i\in l} u^*_{\kk v}(i)\,u_{\kk v'}(i),
\eeq
and the non-Abelian Berry curvature can be evaluated from
\eqs{metric-curvature}{curv-cov-a} with the help of the identity
\beq
\label{eq:A-v}
\ip{u_{\kk c}}{\partial_\kk u_{\kk v}}= -\frac{\me{u_{\kk
      c}}{\hbar\hat\vv_\kk}{u_{\kk v}}}{E_{\kk cv}}.  
\eeq

\subsection{Anisotropic cubic-lattice model}

As our first test case, we consider a model of a ME insulator with no
symmetry. This provides the most challenging case for the theory,
since all nine components of the ME tensor are nonzero and different
from one another.  The resulting surface AHC has both CS and cross-gap
contributions, and it varies with the surface orientation.

We choose the TB model described in Appendix~A of
Ref.~\cite{malashevich-njp10}. This is a spinless model defined on a
cubic lattice, with one orbital per site and eight sites per cell,
where inversion and time-reversal symmetry are broken by assigning
random on-site energies and complex first-neighbor hoppings of fixed
magnitude.  We take the model parameters listed in Table~A.1 of
Ref.~\cite{malashevich-njp10}, and choose the two lowest bands to be
the valence bands. As in that work, all parameters are kept fixed
except for one hopping phase $\varphi$ which is scanned from 0 to
$2\pi$, and the results are plotted as a function of this phase
$\varphi$.

This model is intended as a model for a conventional ME insulator, in
which the isotropic response $\alfa_{\rm iso}/V$ of an insulating
crystallite is very small relative to the quantum $e^2/h$.  The
surface AHC of such a system is most naturally described by setting
$m=0$ in \eqs{sigma-alpha}{ahc-surf-cs} while choosing $\aiso$ and
$\alpha_{\rm CS}$ in the range $[-e^2/2h,e^2/2h]$. (In the next
subsection, we will consider a model with the opposite
characteristics, i.e., with a large isotropic ME response of the order
of $e^2/h$.)

\begin{figure}
  \centering
  \includegraphics[width= 0.95\columnwidth]{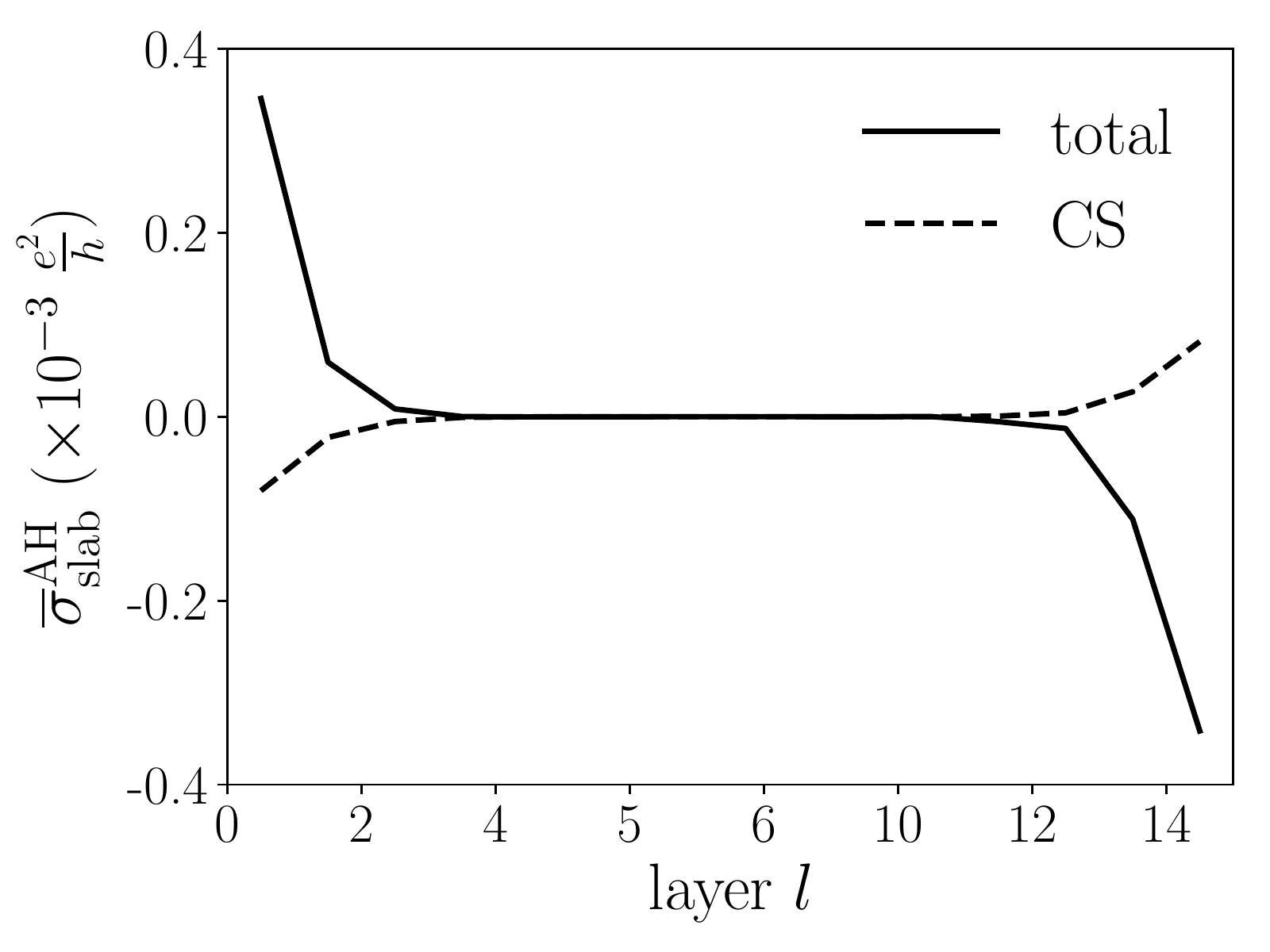}
  \caption{Coarse-grained layer-resolved AHC [\eq{ahc-z-avg-TB}] for a
    16-atom-thick slab of the cubic-lattice model with $\varphi=\pi$.}
  \label{fig:1}
\end{figure}

We construct a slab with a thickness of 16 atomic layers (8 lattice
constants) along~$z$.  The layer-resolved AHC displays strong
oscillations from one layer to the next, which we filter out by
averaging over two consecutive layers:
\beq
\label{eq:ahc-z-avg-TB}
\overline\sigma^{\rm AH}_{\rm slab}(l+1/2)=\frac{1}{2}
\left[
  \sigma^{\rm AH}_{\rm slab}(l)+\sigma^{\rm AH}_{\rm slab}(l+1)
\right].
\eeq
This quantity is plotted as the solid line in \fig{1} for
$\varphi=\pi$, and the dashed line shows the CS contribution the CS
layer-resolved AHC was calculated for a different TB model in
Ref.~\cite{essin-prl09}, and is shown in Fig.~2 therein). Both
quantities are nonzero in the surface regions only, quickly dropping
to almost zero within four subsurface layers.  The fact that the two
curves in \fig{1} are not perfectly odd about the center of the slab
can be attributed to the lack of mirror symmetry in the model. We have
checked that both $\sum_l\,\sigma^{\rm AH}_{\rm slab}(l)$ and
$\sum_l\,\sigma^{\rm AH}_{\rm slab,CS}(l)$ vanish identically, as
should be the case for a slab with $C_{\rm slab}=0$, so that the
macroscopic surface AHC is equal and opposite for the two surfaces.
On a given surface, the CS part of the AHC has the opposite sign
compared to the total. The cross-gap contribution therefore prevails,
as in ordinary ME insulators~\cite{coh-prb11}.

\begin{figure}
  \centering
  \includegraphics[width= 0.95\columnwidth]{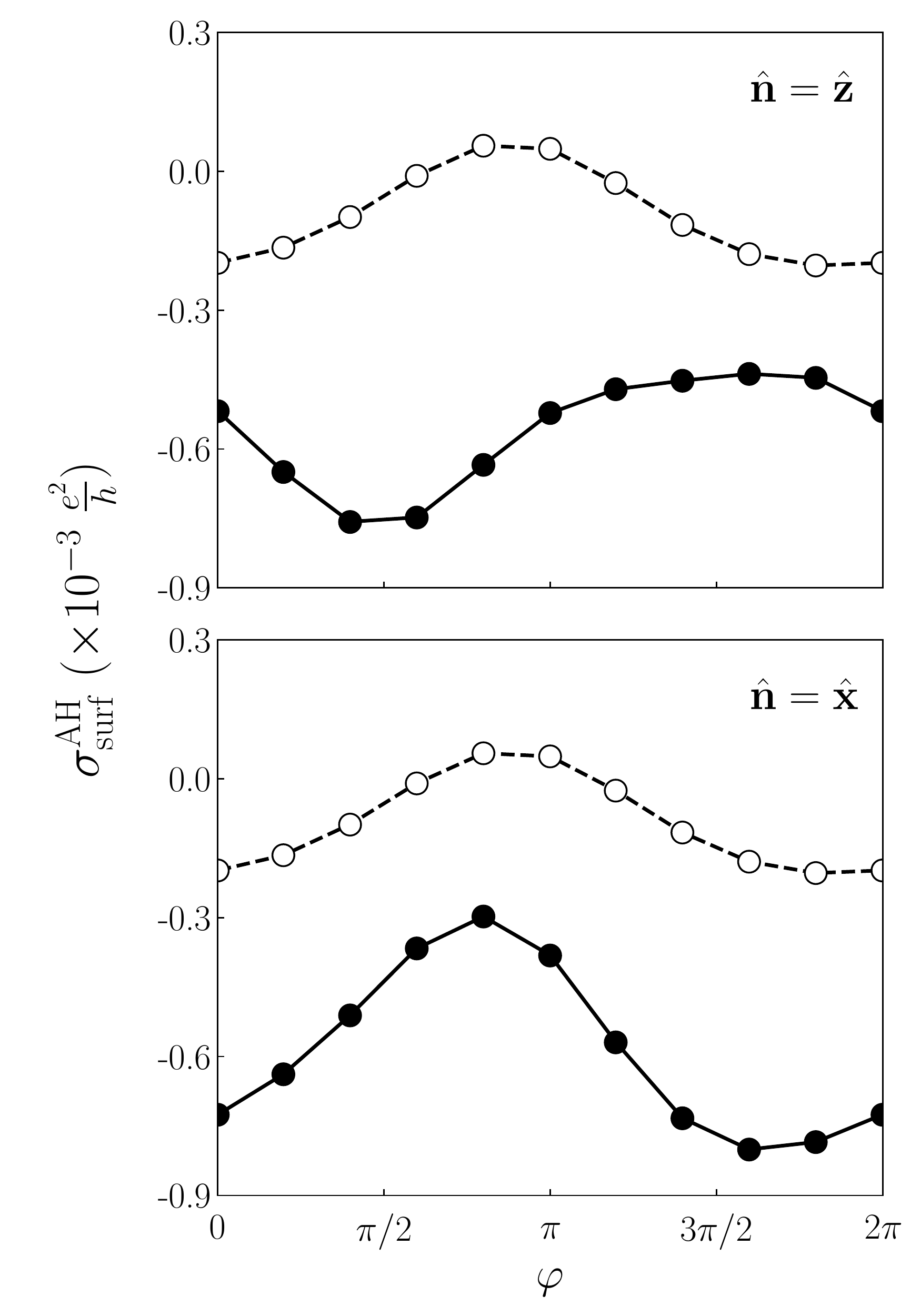}
  \caption{AHC of the top surface (upper panel) and of the right
    surface (lower panel) of 16-layer slabs of the cubic-lattice model
    cut along $z$ and $x$, respectively, as a function of the cyclic
    parameter $\varphi$.  The solid (dashed) lines denote the total
    (CS) surface AHC. Circles represent the quantity appearing on the
    right-hand side of \eq{sigma-alpha-a}, with filled and empty
    circles denoting the total and the CS piece, respectively.}
  \label{fig:2}
\end{figure}

The macroscopic AHC of the top surface is plotted versus $\varphi$ as
the solid line in the top panel of \fig{2}, where the CS contribution
is again shown as a dashed line. For comparison, we plot as filled
(total) and empty (CS) circles the quantities on the right-hand side
of \eqs{sigma-alpha}{ahc-surf-cs}, respectively (with $m=0$). The
precise agreement validates our expression for the surface AHC, and
its decomposition into geometric and nongeometric parts.

In the lower panel of \fig{2} we show results for a slab cut along
$x$. The total surface AHC is different from that on the upper panel,
as expected for an anisotropic model from the last term in
\eq{ahc-surf-cg}. The CS surface AHC is the same in both panels,
confirming that its nonquantized part does not depend on the surface
orientation, as expected from \eq{ahc-surf-cs}.

\subsection{Layered Haldane model}
\label{sec:haldane}

We now turn to a model for which $\aiso$ and $\alpha_{\rm CS}$ are not
always small compared to the quantum $e^2/h$, so that the branch
choice in \eqs{sigma-alpha}{ahc-surf-cs} becomes ambiguous. We choose
the TB model introduced in Ref.~\cite{olsen-prb17b}.  This is a
layered model on a hexagonal lattice, with four orbitals per cell. It
can be obtained by stacking Haldane-model~\cite{haldane-prl88} layers
with alternating parameters, and then coupling them via interlayer
hopping terms.  The model acts as a quantum pump of axion ME coupling:
a slow periodic variation in its parameters can gradually change
$\aiso$ by $e^2/h$ over one cycle. Below, we monitor the evolution of
the surface AHC during one pumping cycle.

We begin by noting that the cross-gap contribution to the ME tensor
vanishes identically for this model.  The reason can be found in
Ref.~\cite{essin-prb10}, where a set of conditions were derived under
which $\alpha^{\rm cg}_{ab}$ vanishes in certain four-band models
having some kind of particle-hole symmetry.  Those conditions hold for
several models proposed in the literature including the present one,
so that the total surface AHC~(\ref{eq:sigma-alpha}) reduces to the CS
part~(\ref{eq:ahc-surf-cs}).

\begin{figure}
  \centering
  \includegraphics[width= 0.95\columnwidth]{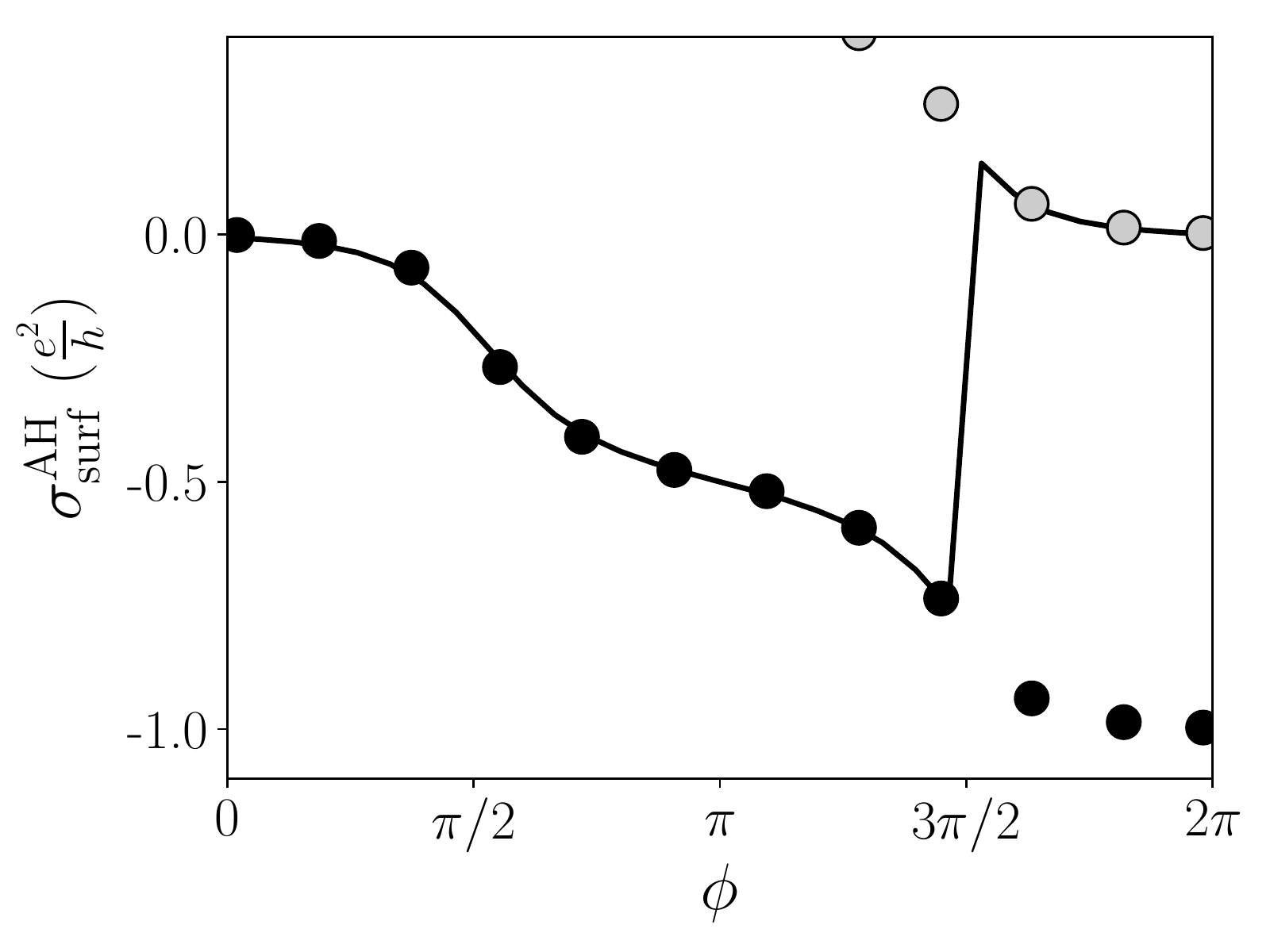}
  \caption{Cyclic evolution of the Hamiltonian of a 10-layer slab of
    the layered Haldane model, during which the isotropic ME coupling
    $\alpha_{\rm CS}$ of the bulk crystal changes by $e^2/h$. The AHC
    of the top surface is plotted as a solid line, and the black and
    gray circles denote two different branches of $-\alpha_{\rm CS}$.
    For a given choice of branch, \eq{ahc-surf-cs} is satisfied
    throughout the cycle with the value of $m$ increasing by one at
    $\phi_c=3\pi/2$.}
  \label{fig:3}
\end{figure}

We construct a slab containing 10 hexagonal layers stacked along~$z$.
The pumping cycle is parametrized by an angle $\phi$ that modulates
the model parameters according to Eqs.~(57c) and~(61) in
Ref.~\cite{olsen-prb17b}.  As in Sec.~III.D of that work, we consider
the situation where the entire slab, including the surfaces, returns
to its initial state at the end of the cycle,
\beq
\label{eq:H-slab-cycle}
\hat H_{\rm slab}(\phi=2\pi)=\hat H_{\rm slab}(\phi=0).
\eeq

The quantities $\sigma^{\rm AH}_{\rm surf}(\phi)$ and
$-\alpha_{\rm CS}(\phi)$ are plotted in \fig{3} as solid lines and
filled circles, respectively; the latter is a multivalued quantity,
and two different branches are shown as black and gray circles.
\equ{ahc-surf-cs} assumes that a specific branch has been chosen, and
we pick the one represented by the black circles. With that choice,
\eq{ahc-surf-cs} is satisfied with $m=0$ for $0<\phi<3\pi/2$ and with
$m=1$ for $3\pi/2<\phi<2\pi$.

The actual value of $m$ at each $\phi$ depends on the particular gauge
choice, but the important point is that \eq{ahc-surf-cs} cannot be
satisfied keeping~$m$ fixed for all $\phi$. \equ{H-slab-cycle} implies
$\sigma^{\rm AH}_{\rm surf}(\phi=2\pi) =\sigma^{\rm AH}_{\rm
  surf}(\phi=0)$,
and the only way this can be reconciled with the pumping of CS axion
coupling in the bulk region,
\beq
\alpha_{\rm CS}(\phi=2\pi)=\alpha_{\rm CS}(\phi=0)+\frac{e^2}{h},
\eeq
is if the integer $m$ in \eq{ahc-surf-cs} increases by one during the
cycle.  This change in the quantized part of the surface AHC is caused
by a topological phase transition at the surface: the energy gap of
the surface bands closes at $\phi_{\rm c}=3\pi/2$, forming a Weyl
point in $(k_x,k_y,\phi)$-space that transfers a quantum of
Berry-curvature flux between the valence and conduction bands as
$\phi$ crosses $\phi_{\rm c}$~\cite{olsen-prb17b}.

\begin{figure}
  \centering
  \includegraphics[width= 0.95\columnwidth]{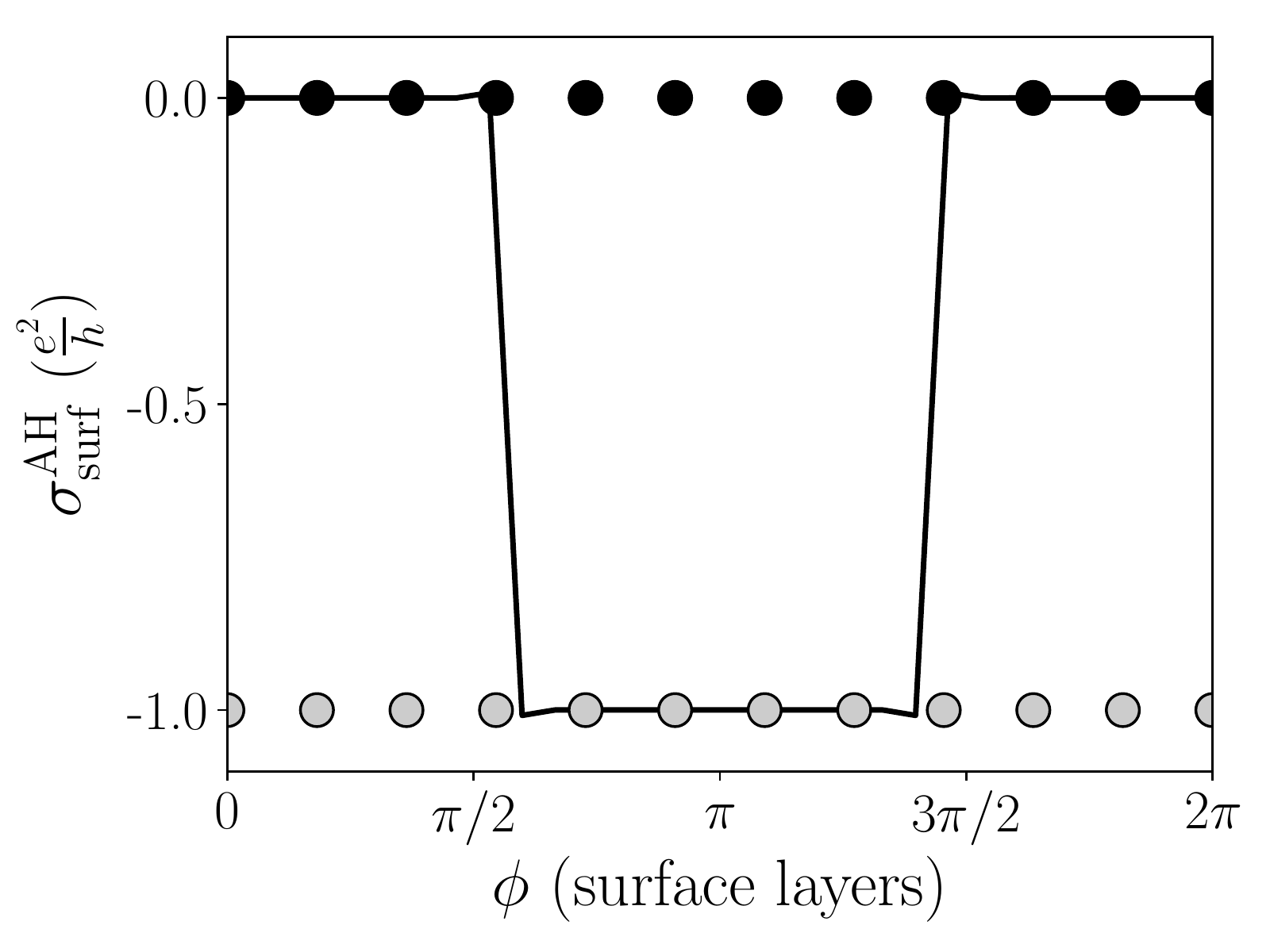}
  \caption{The same quantities as in \fig{3}, but now for a cyclic
    evolution of the Hamiltonian of the surface layers only, keeping
    the Hamiltonian of the rest of the slab fixed.}
  \label{fig:4}
\end{figure}

It is also possible to change the quantized part of the surface AHC by
adjusting only the surface Hamiltonian. This is illustrated in
\fig{4}, where we held the Hamiltonian of all subsurface layers fixed
at $\phi=0$, but modulated the onsite energy of the top and bottom
layers by $\phi$ according to Eq.~(57c) in Ref.~\cite{olsen-prb17b}.
Now the bulk ME coupling $\alpha_{\rm CS}$ is held at zero for all
$\phi$, as indicated by the black circles. The surface AHC vanishes
during half of the cycle leading to $m=0$ in \eq{ahc-surf-cs}, and it
becomes $-e^2/h$ during the other half where $m=-1$. At the critical
points $\phi_c=\pi/2$ and $\phi_c=3\pi/2$, the surface states become
gapless.

\section{Summary}
\label{sec:summary}

In this work we have derived practical expressions for calculating the
full surface AHC of an insulating slab, including the quantized part
that depends on the surface preparation.  That quantized part resides
in a geometric term written in terms of the gauge-covariant Berry
curvature matrix of the slab wave functions. The full surface AHC
contains an additional nongeometric term. Like the nonquantized part
of the geometric surface AHC, that term is only apparently a surface
property, but is in fact fully determined by the bulk ME
tensor~\cite{malashevich-njp10,essin-prb10}.

Numerical TB calculations were carried out to show that our
expressions satisfy the phenomenological relation in \eq{sigma-alpha}
between the surface AHC and the bulk ME coupling. The ability to
change the surface AHC by multiples of $e^2/h$ by adjusting the
surface Hamiltonian was illustrated for a layered Haldane model.

The formalism developed in this work provides a simpler way of
determining the quantized part of the surface AHC than an alternative
approach based on hybrid Wannier functions~\cite{olsen-prb17b}. It
could be particularly useful for characterizing the nontrivial surface
topology in second-order three-dimensional topological insulators with
chiral hinge states~\cite{sitte-prl12,schindler-sciadv18}.

\begin{acknowledgments}
  This work was supported by the Forschungsstipendium Grant No. RA
  3025/1-1 from the Deutsche Forschungsgemeinschaft (T.~R.), by Grant
  No.~FIS2016-77188-P from the Spanish Ministerio de Econom\'ia y
  Competitividad (T.~R. and I.~S.), and by NSF Grant No. DMR-1629059
  (D.~V.).

\end{acknowledgments}

\appendix

\section*{Appendix: Derivation of \eq{ahc-cg-r} for the cross-gap
  local AHC}
\renewcommand{\theequation}{A\arabic{equation}}
\setcounter{equation}{0}

The orbital moment~(\ref{eq:m}) of a finite sample in a static
electric field can be decomposed as~\cite{malashevich-njp10}
%
\begin{subequations}
\label{eq:m-decomp}
\begin{align}
\mm(\EE)&=\mm^{\rm CS}(\EE)+\mm^{\rm cg}(\EE),\\
\mm^{\rm CS}(\EE)&=-\frac{2\pi e^2}{3h}\eps_{ijl}\Im\Tr
\left[ \hat P\hat r_i\hat P\hat r_j\hat P\hat r_l\right]\EE,\\
m^{\rm cg}_{i}(\EE)&=\frac{\pi e}{h}\eps_{ijl}\Im\Tr
\left[ \hat P\hat r_j\hat Q\hat H_0\hat Q\hat r_l -
       \hat Q\hat r_j\hat P\hat H_0\hat P\hat r_l
\right].
\end{align}
\end{subequations}
As in \sect{linres}, $\hat P$ denotes the projection operator onto the
occupied states in the presence of the field, $\hat Q=\one-\hat P$,
and $\hat H_0$ is the unperturbed Hamiltonian. Note that the CS term
has an explicit linear dependence on $\EE$, while the cross-gap (cg)
term only depends on $\EE$ implicitly via the projection operators.
With the above decomposition, the isotropic ME
coupling~(\ref{eq:aiso-tilde}) becomes
\beq
\alfa_{\rm iso}=\alfa^{\rm CS}_{\rm iso}+\alfa^{\rm cg}_{\rm iso},
\eeq
with $\alfa^{\rm CS}_{\rm iso}$ given by \eq{alpha-iso-cs} and
\beq
\label{eq:alfa-cg-iso}
\alfa^{\rm cg}_{\rm iso}=-\frac{1}{3}\int r_i
\eps_{ijl}
\left.\frac{\partial T_j(\rr,\EE)}{\partial\E_l}\right|_{\EE=0}\,d^3r,
\eeq
where 
\beq
\label{eq:T}
T_j(\rr,\EE)=\frac{\pi e}{h}\Im
\me{\rr}{\hat P\hat r_j\hat Q\hat H_0\hat Q -
         \hat Q\hat r_j\hat P\hat H_0\hat P}{\rr}.
\eeq
Comparing \eq{alfa-cg-iso} with \eq{alfa-sigma} for $\alfa_{\rm iso}$
we conclude that the cross-gap local AHC is given by
\beq
\label{eq:ahc-cg-r-a}
\sigma^{\rm AH}_{{\rm cg},i}(\rr)=
\eps_{ijl}
\left.\frac{\partial T_j(\rr,\EE)}{\partial\E_l}\right|_{\EE=0}.
\eeq
Our remaining task is to show that this expression is equivalent to
\eq{ahc-cg-r}.

We begin by plugging \eq{T} into \eq{ahc-cg-r-a}. This generates a
total of six terms,
\begin{widetext}
\bea
\label{eq:ahc-cg-r-b}
\sigma^{\rm AH}_{{\rm cg},i}(\rr)=\frac{\pi e}{h}\eps_{ijl}\Im
\bra{\rr}
&\Big[&
  -\hat P_0\hat r_j\left(\hat H_0\hat Q_0\hat P_l\right)
  -\hat P_0 \hat r_j \left(\hat P_l\hat Q_0\hat H_0\right)
  +\hat P_l\hat r_j\hat H_0\hat Q_0\nn
  &&-\hat Q_0\hat r_j\left(\hat H_0\hat P_0\hat P_l\right)
  -\hat Q_0\hat r_j\left(\hat P_l\hat P_0\hat H_0\right)
  +\hat P_l\hat r_j\hat H_0\hat P_0
\Big]
\ket{\rr},
\eea
\end{widetext}
where we used the fact that $\hat H_0$ commutes with both $\hat P_0$
and $\hat Q_0$, and introduced the notation
$\hat P_l=\partial_{\E_l}\hat P=-\partial_{\E_l}\hat Q$ for the
Cartesian components of $\partial_\EE\hat P$.  Using \eq{dPdE} for
that operator, the individual terms in \eq{ahc-cg-r-b} become
\begin{widetext}
\begin{subequations}
\label{eq:terms}
\begin{align}
\label{eq:term1}
-\Im\me{\rr}{\hat P_0\hat r_j\left(\hat H_0\hat Q_0\hat P_l\right)}{\rr}
&=\frac{eE_c}{E_{cv}}\Im
\left[
  \ip{\rr}{v'}\me{v'}{\hat r_j}{c}\me{c}{\hat r_l}{v}\ip{v}{\rr}
\right],\\
\label{eq:term2}
-\Im\me{\rr}{\hat P_0 \hat r_j 
\left(\hat P_l\hat Q_0\hat H_0\right)}{\rr}
&=\frac{eE_c}{E_{cv}}\Im
\left[
  \ip{\rr}{v'}\me{v'}{\hat r_j}{v}\me{v}{\hat r_l}{c}\ip{c}{\rr}
\right],\\
\label{eq:term3}
\Im\me{\rr}{\hat P_l\hat r_j\hat H_0\hat Q_0}{\rr}
&=-\frac{eE_{c'}}{E_{cv}}\Im
\left[
  \ip{\rr}{c}\me{c}{\hat r_l}{v}\me{v}{\hat r_j}{c'}\ip{c'}{\rr}
\right]
-\frac{eE_{c'}}{E_{cv}}\Im
\left[
  \ip{\rr}{v}\me{v}{\hat r_l}{c}\me{c}{\hat r_j}{c'}\ip{c'}{\rr}
\right],\\
\label{eq:term4}
-\Im\me{\rr}{\hat Q_0\hat r_j\left(\hat H_0\hat P_0\hat P_l\right)}{\rr}
&=\frac{eE_v}{E_{cv}}\Im
\left[
  \ip{\rr}{c'}\me{c'}{\hat r_j}{v}\me{v}{\hat r_l}{c}\ip{c}{\rr}
\right],\\
\label{eq:term5}
-\Im\me{\rr}{\hat Q_0\hat r_j\left(\hat P_l\hat P_0\hat H_0\right)}{\rr}
&=\frac{eE_v}{E_{cv}}\Im
\left[
  \ip{\rr}{c'}\me{c'}{\hat r_j}{c}\me{c}{\hat r_l}{v}\ip{v}{\rr}
\right],\\
\label{eq:term6}
\Im\me{\rr}{\hat P_l\hat r_j\hat H_0\hat P_0}{\rr}
&=-\frac{eE_{v'}}{E_{cv}}\Im
\left[
  \ip{\rr}{c}\me{c}{\hat r_l}{v}\me{v}{\hat r_j}{v'}\ip{v'}{\rr}
\right]
-\frac{eE_{v'}}{E_{cv}}\Im
\left[
  \ip{\rr}{v}\me{v}{\hat r_l}{c}\me{c}{\hat r_j}{v'}\ip{v'}{\rr}
\right],
\end{align}
\end{subequations}
\end{widetext}
where a summation over repeated band indices is implied.  We wish to
bring the sum of all these terms into a ``cross-gap'' form, where
dipole matrix elements only connect occupied and empty states. Four of
the eight terms above already have that form and they can be combined
in pairs, (\ref{eq:term1}) with the second term in (\ref{eq:term6})
and the first term in (\ref{eq:term3}) with (\ref{eq:term4}), to get
\bea
&&\frac{e(E_c+E_{v'})}{E_{cv}}\Im
\left[
  \ip{\rr}{v'}\me{v'}{\hat r_j}{c}\me{c}{\hat r_l}{v}\ip{v}{\rr}
\right]\nn
&+&\frac{e(E_v+E_{c'})}{E_{cv}}\Im
\left[
  \ip{\rr}{c'}\me{c'}{\hat r_j}{v}\me{v}{\hat r_l}{c}\ip{c}{\rr}
\right].
\eea

In the remaining four terms, we use the completeness relation to bring
them to the desired form.  First we replace $\ket{v'}\bra{v'}$ with
$\one-\ket{c'}\bra{c'}$ in (\ref{eq:term2}) and $\ket{c'}\bra{c'}$
with $\one-\ket{v'}\bra{v'}$ in (\ref{eq:term5}).  The two terms
containing $\one$ can be reduced to
\beq
\label{eq:zero}
-r_j\Im
\left[
  \ip{\rr}{c'}\me{c'}{\hat r_l}{c}\ip{c}{\rr}
\right]=0,
\eeq
leaving
\bea
&-&\frac{eE_c}{E_{cv}}\Im
\left[
  \ip{\rr}{c'}\me{c'}{\hat r_j}{v}\me{v}{\hat r_l}{c}\ip{c}{\rr}
\right]\nn
&-&\frac{eE_v}{E_{cv}}\Im
\left[
  \ip{\rr}{v'}\me{v'}{\hat r_j}{c}\me{c}{\hat r_l}{v}\ip{v}{\rr}
\right].
\eea
\\
Next we combine the second term in (\ref{eq:term3}) with the first in
(\ref{eq:term6}) using
$\hat H_0=E_{v'}\ket{v'}\bra{v'}+E_{c'}\ket{c'}\bra{c'}$,
\begin{widetext}
\bea
&&\frac{e}{E_{cv}}\Im
\left[
  \me{\rr}{\hat H_0\hat r_j}{v}\me{v}{\hat r_l}{c}\ip{c}{\rr}+
  \me{\rr}{\hat H_0\hat r_j}{c}\me{c}{\hat r_l}{v}\ip{v}{\rr}
\right]\nn
&-&\frac{e}{E_{cv}}\Im
\left[
  E_{c'}\ip{\rr}{c'}\me{c'}{\hat r_j}{v}\me{v}{\hat r_l}{c}\ip{c}{\rr}+
  E_{v'}\ip{\rr}{v'}\me{v'}{\hat r_j}{c}\me{c}{\hat r_l}{v}\ip{v}{\rr}
\right].
\eea
\end{widetext}
Writing $\hat H_0\hat r_j$ as $\hat r_j\hat H_0-i\hbar\hat v_j$ and
then canceling two terms according to \eq{zero}, the first line
becomes
\beq
-\frac{e\hbar}{E_{cv}}\Re
\left[
  \me{\rr}{\hat v_j}{v}\me{v}{\hat r_l}{c}\ip{c}{\rr}+
  \me{\rr}{\hat v_j}{c}\me{c}{\hat r_l}{v}\ip{v}{\rr}
\right].
\eeq
Collecting terms in \eq{ahc-cg-r-b} for the cross-gap local AHC we
find, after some cancellations,
\begin{widetext}
\bea
\label{eq:sigma-kubo-r-final}
\sigma^{\rm AH}_{{\rm cg},i}(\rr)
=&-&\frac{e^2}{2E_{cv}}\eps_{ijl}\Re
\left[
  \me{\rr}{\hat v_j}{v}\me{v}{\hat r_l}{c}\ip{c}{\rr}+
  \me{\rr}{\hat v_j}{c}\me{c}{\hat r_l}{v}\ip{v}{\rr}
\right]\nn
&+&\frac{\pi e^2}{h}\eps_{ijl}\Im
\left[
    \ip{\rr}{v'}\me{v'}{\hat r_j}{c}\me{c}{\hat r_l}{v}\ip{v}{\rr}-
    \ip{\rr}{c'}\me{c'}{\hat r_j}{v}\me{v}{\hat r_l}{c}\ip{c}{\rr}
\right].
\eea
\end{widetext}
Comparing the first line with \eq{dPdE} for $\partial_\EE\hat P$ and
using projection operators in the second line, we arrive at
\begin{widetext}
\bea
{\bm \sigma}^{\rm AH}_{{\rm cg}}(\rr)
=\frac{e}{2}\,\Re
  \bra{\rr}\hat \vv\times\partial_\EE\hat P\ket{\rr}
+\frac{\pi e^2}{h}\Im
\left[
  \me{\rr}{\hat P_0\hat\rr\hat Q_0\times\hat Q_0\hat\rr\hat P_0}{\rr}-
  \me{\rr}{\hat Q_0\hat\rr\hat P_0\times\hat P_0\hat\rr\hat Q_0}{\rr}
\right].
\eea
\end{widetext}
As noted earlier,\textsuperscript{\ref{foot:PHC}} the last two terms
in this expression are equal to one another, resulting in
\eq{ahc-cg-r}.


%

\end{document}